\documentclass[prd,preprint,showpacs,preprintnumbers,floats,floatfix,apsrev,amsmath,amsfonts,nofootinbib,superscriptaddress]{revtex4-1}
\usepackage{color}
\usepackage{graphicx}
\usepackage{dcolumn}
\usepackage{natbib}
\usepackage{hyperref}
\usepackage{hypernat}

\newcommand{\di}{\mathrm{d}}

\newcommand{\be}{\begin{equation}}
\newcommand{\en}{\end{equation}}

\begin{document}
\title{Epidemiological projections for COVID-19 considering lockdown policies and social behavior: the case of Bolivia}
\author{M.~L.~Pe\~{n}afiel}
\email{mpenafiel@cbpf.br}
\affiliation{CBPF - Brazilian Center for Research in Physics, Xavier Sigaud st. 150, zip 22290-180, Rio de Janeiro, RJ, Brazil.}

\author{G.~M.~Ram\'{i}rez-\'{A}vila}
\email{gramirez@ulb.ac.be}
\affiliation{Instituto de Investigaciones F\'{i}sicas, Universidad Mayor de San Andr\'{e}s, Casilla 8635, La Paz, Bolivia.}
\date{\today}            

\begin{abstract}
We assess the epidemic situation caused by SARS-CoV-2 using Tsallis' proposal for determining the occurrence of the peak, and also the Susceptible-Infected-Recovered-Asymptomatic-Symptomatic and Dead (\textbf{SIRASD}) compartmental model. Using these two models, we determine a range of probable peak dates and study several social distancing scenarios during the epidemic. Due to the socioeconomic situation and the conflictive political climate, we take for our study the case of Bolivia, where a national election was originally scheduled to occur on September 6th and recently rescheduled to be on October 18th. For this, we analyze both electoral scenarios and show that such an event can largely affect the epidemic's dynamics.
\end{abstract}

\maketitle

\section{Introduction}

COVID-19 pandemic has shook the world. Since the official notification of the detection of the novel corona-virus, a huge scientific effort has been made all over the world in many respects, e.g. the investigation of a prospective treatment or vaccine \cite{Shih2020,Alsuliman2020}, the proper mechanisms of the virus and new epidemiological models for the disease spreading \cite{Ndaierou2020}, among others. On the other hand, the pandemic control around the world has tested the sanitary and economic policies of almost any country in the world, leading most of the countries to impose lockdown periods \cite{Mitja2020}. Underdeveloped or developing countries represent a particularly dangerous scenario for the spreading of COVID-19, since both scientific and governmental efforts are strongly limited due to the lack of economic and human resources. 

Latin America has become the new epicentre for the spreading of COVID-19 during the past month, with specially concerning cases in Brazil, Mexico, Ecuador, Per\'{u} and Chile, with an overall of over 3 million cases as of July, 2020. Bolivia, in its own right, is a particularly concerning country due to the high percentage of population living on a basis of informal economy\footnote{Bolivian informal economy reached 62.3\% of the Gross Domestic Product (GDP) as of 2018 according to the International Monetary Fund (IMF) \cite{Medina2018}.}, recent political events leading to exacerbated polarization, and a deficient healthcare system. Moreover, a national election was originally scheduled for September 6, this kind of event implies the mobilization of over 7 million people across the country (representing about 70\% of whole population). Furthermore, the epidemiological peak for Bolivia is expected to occur in late August or early September according to the public health authorities\footnote{Nevertheless, no public data about these projections is available.}, meaning that the national election was going to take place in the worst possible scenario: the epidemiological peak. Based on their predictions; recently, the local electoral authorities have re-scheduled the  poll day to occur on October 18.

Across Latin America, several local efforts (see \cite{FiescoSepulveda2020,CaicedoOchoa2020} and the references therein) have been made in order to predict and forecast the epidemiological peaks in each country (c.f. \cite{Crokidakis2020,Tsallis2020}) taking into account, for instance, several political, economical \cite{Pires2020} and social measures \cite{Bastos2020a}. These works are of particular importance in the actual crisis due to the need of governments for taking the \emph{less harmful} actions in order to overcome the current scenario.

Specifically, the current situation (as of July 2020) in Bolivia is alarming. Due to the economic urgency, the rigid quarantine was lifted on June 1 in most of the country, and data suggests that social distancing measures were technically loosen weeks before the official release date. Moreover, both public and private healthcare systems are collapsed and the national diagnosing facilities had to overcome several human and material deficiencies, conducting to a publicly recognized sub-estimation of both active and recovered cases. Also, the lack of an effective treatment has led people to search for miraculous cures, such as the ``miracle mineral solution'' (MMS) or chlorine dioxide solution (CDS), among others \cite{Zhou2020}, guided by several highly influential journalists and some local authorities and medicine practitioners, which might eventually lead to a overcharging of the healthcare system not only due to COVID-19.

In this work we intend to tackle the problem of the forthcoming national election from a statistical point of view, by analyzing the situation of a massive mobilization given the current projections based on reliable epidemiological models. We also intend to address the problem of imposing a \emph{rigid} quarantine before and after the epidemic peak, assessing the potential case reduction and appearance of a second peak following the first one. For this respect, in Sec. \ref{sec:fit} we use a fit model proposed by Tsallis and Tirnakli \cite{Tsallis2020} with the purpose of fitting the Bolivian data for active cases and estimate a time range for the peak date of the epidemic curve. In Sec. \ref{sec:model} we review the \textbf{SIRASD} compartmental model which accounts for a fraction of the infected population being asymptomatic, which, as is known, is the case for COVID-19. Furthermore, we estimate the epidemiological parameters for the early evolution of the epidemic in Bolivia and propose several scenarios for the evolution of the epidemic curves. In Sec. \ref{sec:results} we discuss the results obtained for the different epidemic scenarios proposed; finally in Sec. \ref{sec:conclusions} we conclude and offer perspectives for further developments.

We emphasize that the intention of the present work \textbf{does not} consist in predicting a date for the occurrence of the epidemiological peak, but rather on the analysis of the social distancing policies near the epidemic peak and near an extraordinary event implying a massive concentration of persons that might dramatically heighten the contagion: the election day.

\section{Fitting the Bolivian data} \label{sec:fit}
The first case of COVID-19 in Bolivia was diagnosed on March 9th \cite{EscaleraAntezana2020} and, up to July 12th the local health authorities have accounted for 47200 positive cases. Tsallis and Tirnakli \cite{Tsallis2020} inspired on a stock-market model proposed an analytic function that fits several \emph{full} epidemiological curves for countries such as China, South Korea, France, etc. with the aim of forecasting their epidemiological curves. They have proposed the following functional form for the behavior of the COVID-19 active cases
\be \label{eq:Tsallis}
N=C\frac{(x-x_0)^{\alpha}}{(1+(q-1)\bar{\beta} (x-x_0)^{\bar{\gamma}})^{1/(1-q)}}\ ,
\en
where $C>0,\ \alpha>0,\ \bar{\beta}>0,\ \bar{\gamma}>0,\ q>0$ and $x$ accounts for the time elapsed since the first case measured in days. Eq.~ \eqref{eq:Tsallis} is known as a \emph{q-exponential} function. The parameters in \eqref{eq:Tsallis} are interpreted as $(C,x_0)$ being trivial parameters depending on the country's population and initial day of the pandemic with respect to China's data\footnote{Since $x_0$ represents a shift in the origin for the beginning of the country's epidemic, we will take $x_0=0$ for sake of simplicity.}, respectively; $(q,\bar{\gamma})$ being disease-specific parameters that may account for the biological aspects of COVID-19 and are quite universal in the many cases analyzed, with $q\approx1.26$ and $\bar{\gamma}\approx3$ \cite{Tsallis2020}. Finally, $(\alpha,\bar{\beta})$ account for the country's particular spreading of the disease, which include social distancing policies, the adequate control of the isolation of infected persons and might, as well, involve particular environmental conditions that could alter the disease propagation and its severity \cite{AriasReyes2020,Soliz2020}.

\begin{figure}[b]
\includegraphics[width=0.77\textwidth]{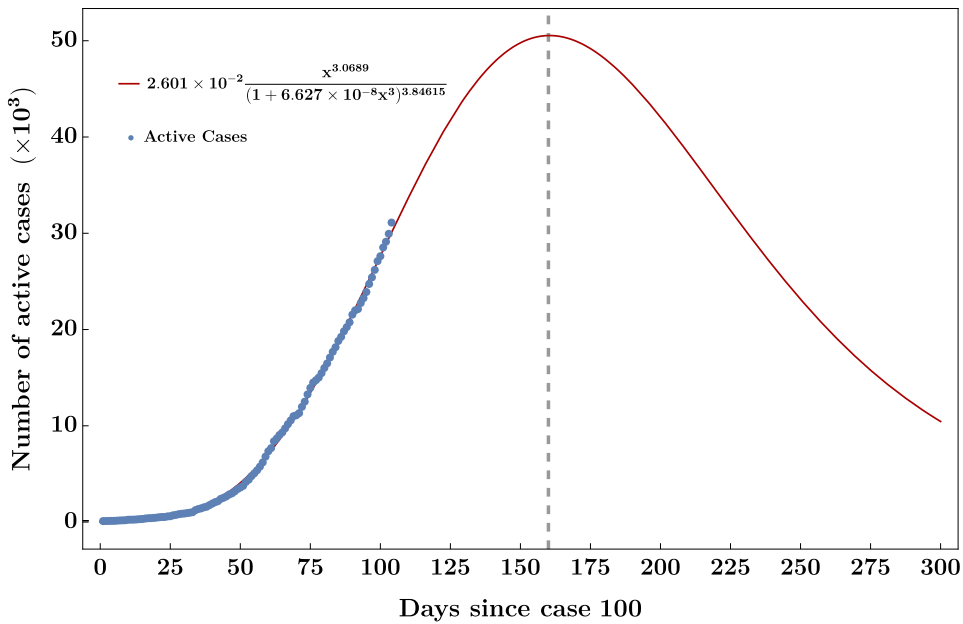}
\caption{Data for the active cases in Bolivia as of July 11th, the fit corresponds to the function proposed in \cite{Tsallis2020}. According to this fit, the peak of infections is expected to occur on the day 160, i.e. September 5th.}
\label{fig:Tsallisfit}
\end{figure}

\begin{figure}[b]
\includegraphics[width=0.76\textwidth]{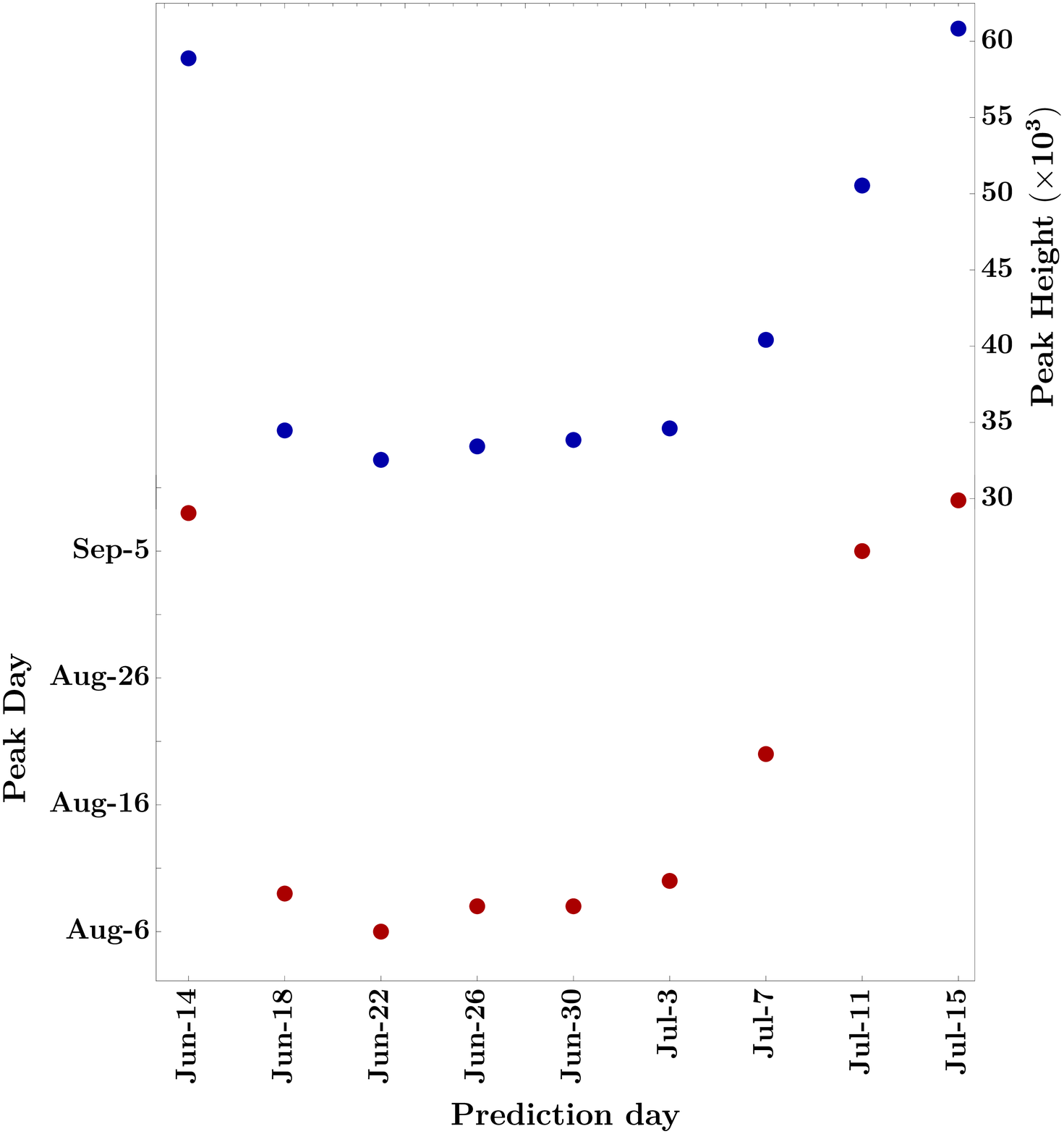}
\caption{Projections for the peak date and height of active cases corresponding to data available on successive dates according to Tsallis' fit.}
\label{fig:peakgraph}
\end{figure}

Figure \ref{fig:Tsallisfit} shows the active cases data fitted with Eq.~ \eqref{eq:Tsallis}, the relevant parameters for this fit are $C=2.601\times10^{-2}$, $\alpha=3.0689$, $\bar{\beta}=2.5492\times10^{-7}$ while the pair $(\bar{\gamma},q)=(3,1.26)$ is fixed to the Chinese parameters since the local epidemic in Bolivia has not reached its peak yet. We observe that Fig.~ \ref{fig:Tsallisfit} is in good agreement with the available data. It is clear that for the parameter fitting, its quality is improved with the quantity of available data. Interestingly enough, we can perform such a parameter estimation for available data at different dates and calculate both the epidemic peak date and its height. Figure \ref{fig:peakgraph} shows this calculation for different available data. According to the fit function, \eqref{eq:Tsallis}, for the available data proposed in \cite{Tsallis2020}, the Bolivian epidemic peak is expected to occur in the range [Aug-6,Sep-6], which coincides with local authorities' information divulgation.

\section{Model} \label{sec:model}
The compartmental epidemiological model we use in the present work is the so-called Susceptible-Infected-Recovered for Asymptomatic-Symptomatic and Dead (\textbf{SIRASD}) model \cite{Bastos2020} modified in order to account for the sorting of the total population in two groups \cite{Pires2020}, those who are not economically obliged to break social distancing measures (i.e., the fraction of the population possessing a \emph{formal} work), and those who need to break social distancing measures due to economical reasons (i.e., the fraction of the population living in the \emph{informal} sector of the economy). This model describes the evolution of a disease accounting for the infected population consisting of asymptomatic ($A_j$) and symptomatic (infected) individuals ($I_j$) in each group; which, as is known, is the case for the COVID-19 pandemic. For sake of simplicity, only the symptomatic fraction of the infected individuals is susceptible to die from the disease.The model is given by the following equations
\begin{subequations} \label{eq:SIRASD}
\begin{eqnarray}
\frac{\di S_j}{\di t}&=&-\sum_{k=1}^{2}\phi_j\phi_k\left(\beta_A A_k+\beta_I I_k\right)\frac{S_j}{N}\ , \\
\frac{\di A_j}{\di t}&=&\left(1-p\right)\sum_{k=1}^{2}\phi_j\phi_k\left(\beta_A A_k+\beta_I I_k\right)\frac{S_j}{N}-\gamma_A A_j\ , \\
\frac{\di I_j}{\di t}&=&p\sum_{k=1}^{2}\phi_j\phi_k\left(\beta_A A_k+\beta_I I_k\right)\frac{S_j}{N}-\gamma_I I_j\ , \\
\frac{\di R_j}{\di t}&=&(1-r)\gamma_I I_j+\gamma_A A_j\ , \\
\frac{\di D_j}{\di t}&=&r\gamma_I I_j\ , \\
\frac{\di N_j}{\di t}&=&-2r\gamma_I I_j\ ,
\end{eqnarray}
\end{subequations}
where $S_j$ accounts for the susceptible fraction of the population, $A_j$ for the asymptomatic fraction, $I_j$ for the symptomatic fraction (which is more likely to be tested positive), $R_j$ for the recovered fraction of the total infected individuals and $D_j$ for the fraction of the population that dies from the disease, the sub-index $j$ takes account of the group label; consequently, for our case it can be either $1$ or $2$. Since this model assumes a portion of the population dies from the disease, the total number of individuals is not constant over time and is given by $N(t)=N_1(t)+N_2(t)=\sum_{j=1}^{2}\left(S_j+I_j+A_j+R_j-D_j\right)$. Furthermore, the parameter $\phi_j$ represents the noncompliance degree of the social distancing measures corresponding to the group $j$ \cite{Pires2020}. Note that Eqs.~ \eqref{eq:SIRASD} take account of intragroup interactions ($\phi_j\phi_j$) as well as interactions between members of different groups ($\phi_j\phi_k, \ j\ne k$) 

For the purpose of estimating the parameters of the model we first need to use the simpler \textbf{SIRD} model \cite{Bastos2020}, which does not assume an asymptomatic fraction of the population and consists of the population composed by only one group. This model is given by the same equations as \eqref{eq:SIRASD} with the detail that the compartments $A_j$ are removed, as well as their corresponding parameters $\beta_A$ and $\gamma_A$ \cite{Bastos2020}.

The first 2 COVID-19 cases in Bolivia were diagnosed on March 9, 2020. Soon after, the central government took several social distancing policies, such as the closure of educational establishments at all levels and a partial quarantine, that consisted on a reduction of the working hours of public and private institutions, including the suspension of cultural, religious and sporting events; however, these measures did not impose, in practice, limits for potential crowds to gather, e.g.,  public transportation and political manifestations, among others. These measures lasted for 13 days until March 22, where the government announced a rigid quarantine, allowing citizens to stay out of their homes 1 day per week and banning both public and private transportation. Assuming that this final measure is reflected in the data with a delay of $\sim2$ weeks after the measure, we will estimate the epidemiological parameters using the first 27 days of epidemiological data.

\begin{figure}[t]
\includegraphics[width=0.76\textwidth]{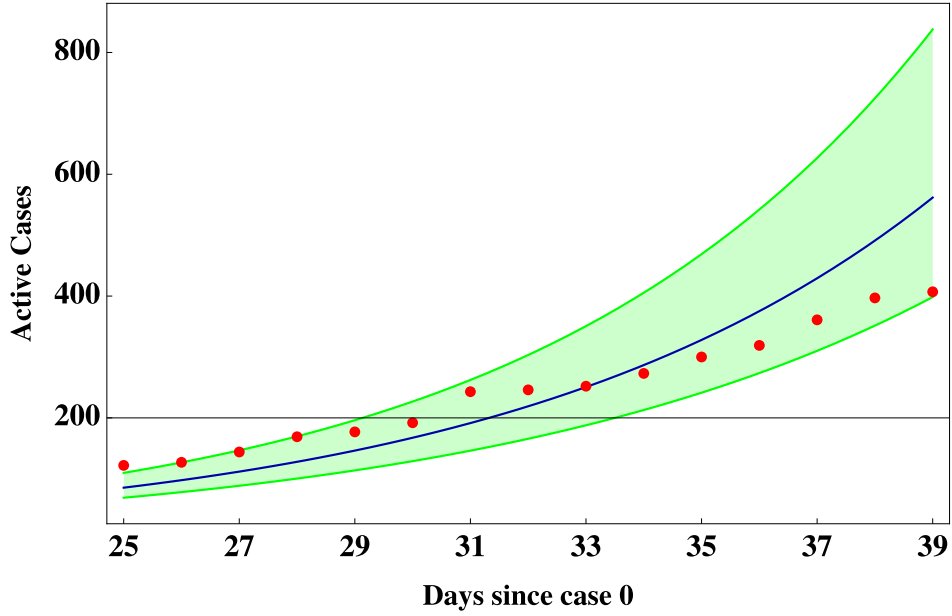}
\caption{ Early evolution for infected individuals for the parameters estimated for the \textbf{SIRASD} model assuming one population only. The red points are the real data for infected cases, the blue line is the estimated infection curve and the green region represents the 95\% confidence interval for the estimated parameters.}
\label{fig:earlyp95}
\end{figure}

In order to estimate the parameters it is necessary to solve the system of nonlinear differential equations (Eq.~\eqref{eq:SIRASD} for the case of the \textbf{SIRASD} model) and minimize the square error with respect to the relevant parameters \cite{Bastos2020}. Hence, we estimate the relevant parameters in the early stage of the epidemics using the \textbf{SIRD} model as $\beta_I\sim0.457868$, $\gamma_I\sim0.345233$, $r\sim0.0677$. By replacing these parameters in Eq.~ \eqref{eq:SIRASD} we can further estimate the remaining parameters as $\beta_A\sim0.437923,\ \gamma_A\sim0,280624,\ p\sim0.544546$. Figure~\ref{fig:earlyp95} shows the infections curve for the early evolution of the epidemic for the single population \textbf{SIRASD} model using these parameters, additionally we plot the 95\% confidence interval for these parameters. Since the scope of the work is to give a qualitative description of social distancing measures, for sake of clarity, the rest of the epidemiological curves are plotted only using the estimated parameters.

\begin{figure}[t]
\includegraphics[width=0.76\textwidth]{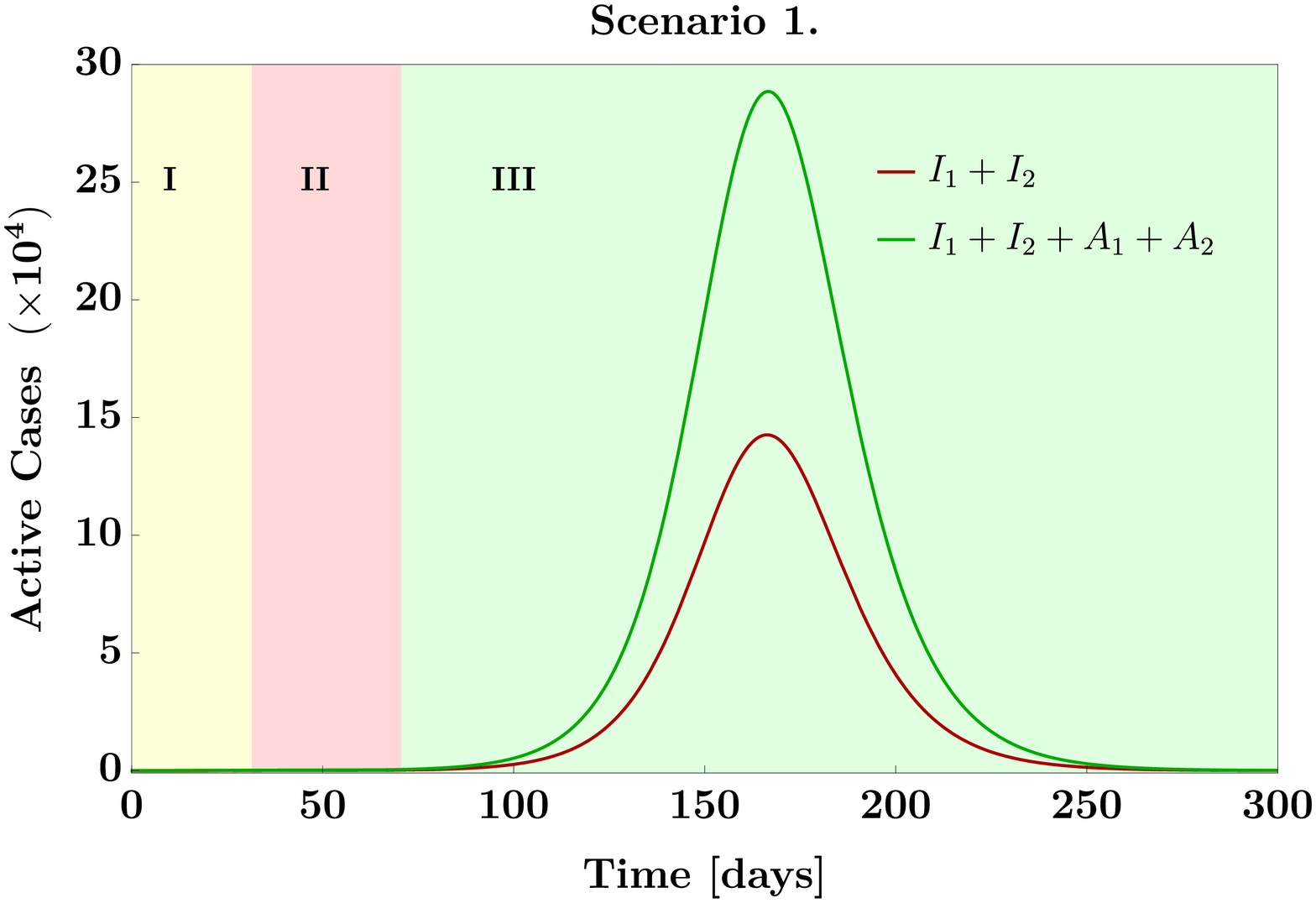}
\caption{ Scenario 1. Region \textbf{I} corresponds to $\phi_1=\phi_2=1$, region \textbf{II} corresponds to $\phi_1=0.8,\ \phi_2=0.9$ and region \textbf{III} to $\phi_1=0.87, \phi_2=0.98$.}
\label{fig:scen12}
\end{figure}

With the intention of assessing the behavior of the epidemic in Bolivia we will explore three scenarios:
\begin{enumerate}
\item \label{sce:1} The free evolution of the epidemiological curves, where region \textbf{I} encompasses the first 27 days of the epidemic and $\phi_1=\phi_2=1$; region \textbf{II} covers the rigid quarantine period, from day 27 to day 67 and we set $\phi_1=0.8$, $\phi_2=0.9$. Finally, region \textbf{III} encompasses the rest of the epidemic with lax social distancing measures and $\phi_1=0.87$, $\phi_2=0.98$.
\item \label{sce:3}The scenario where rigid social distancing measures or lockdown are imposed near the epidemiological peak and the social distancing measures are lifted at a given time $t_\text{off}$ after the peak, as in \cite{Pires2020}. Regions \textbf{I} and \textbf{II} are the same as in scenario \ref{sce:1}, while region \textbf{III} ends at $t_\text{in}=t_{peak}-n,\ n\in\mathbb{N}$ with $t_{peak}=166$. Region \textbf{IV} starts at $t_\text{in}$ and ends at $t_\text{off}$ with $t_\text{in}$ and $t_\text{off}$ having variable values. Finally, region \textbf{V} goes from $t_\text{off}$ to the end of the epidemic. The values of $\phi_j$ for the first three regions are the same as in scenario \ref{sce:1}, while for region \textbf{IV} we set $\phi_1=0.85,\ \phi_2=0.94$ and region \textbf{V} has $\phi_1=0.9,\ \phi_2=0.95$.
\item \label{sce:5} The scenario considering that the poll day occurs on September 6th, which corresponds to day 181 (and on October 18th, day 223), and no social distancing policies are adopted during the peak. Such a democratic event implies the massive mobilization of people not only on the election day but also on periods both before and after the event. For this respect we will assume that massive mobilizations ($\phi_1=\phi_2=1$) take place one week before and after the event. For this scenario we have regions \textbf{I} and \textbf{II} unaltered and region \textbf{III} goes up to day 174 for September (216 for October); for region \textbf{IV} we set $\phi=1$ and it goes up to day 188 for September (230 for October). Finally, region \textbf{V} recovers the values of \textbf{III} for $\phi$ and lasts until the end of the evolution.
\end{enumerate}
\section{Results} \label{sec:results}
We set Bolivia's population to be $N_0=11677580$. In order to solve the set of Eqs.~ \eqref{eq:SIRASD} we consider the initial conditions $S_0^{(j)}=f_jN_0-A_j(0)-I_j(0)-R_j(0)-D_j(0)$ according to the first reported cases informed by the Ministry of Health. Thus, we set $I_1(0)=2$, $A_1(0)=2$ and the rest of the initial conditions to 0. $f_j$ is the fraction of the total population which belongs to each social group $j$; therefore we have $1=f_1+f_2$. For the Bolivian case, we set $f_1=0.38$ and $f_2=0.62$ assuming that group 2 consists exclusively on the fraction of the population living on a basis of informal economy according to the IMF data \cite{Medina2018}.

\begin{figure*}[b]
\includegraphics[width=0.46\textwidth]{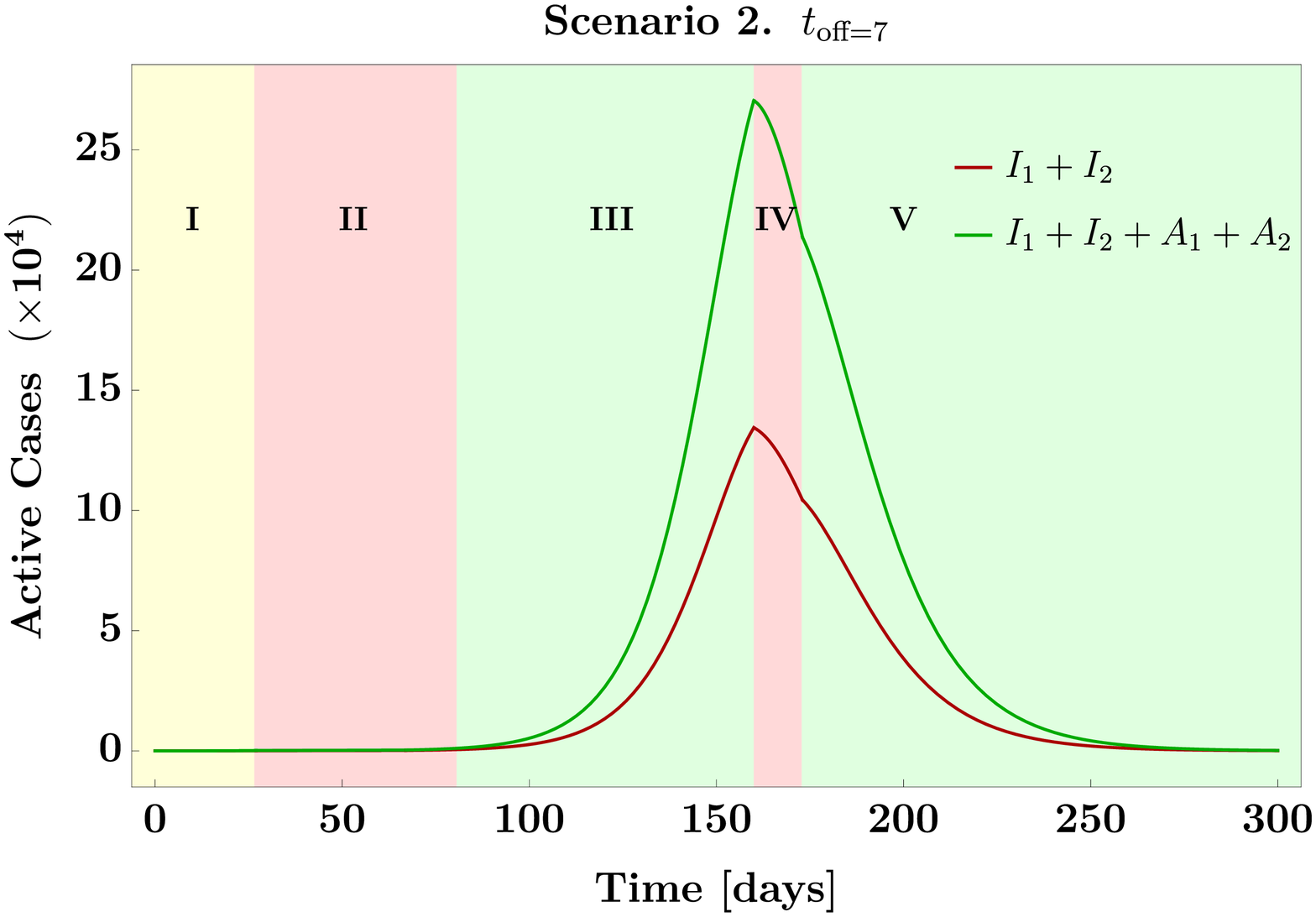}\qquad
\includegraphics[width=0.46\textwidth]{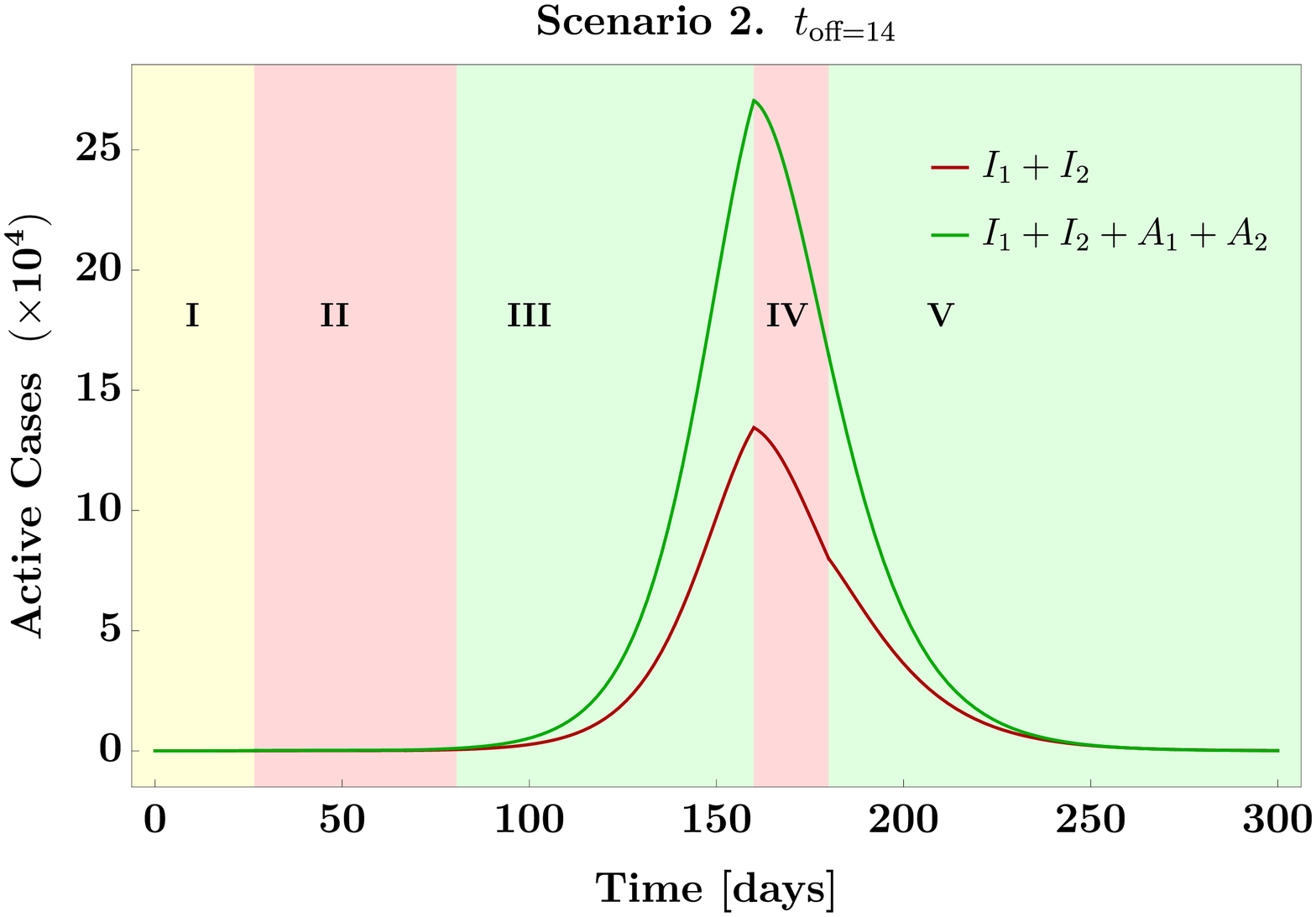}
\caption{(Left) Scenario 3 with $t_\text{off}=7$. For the parameters chosen for region \textbf{IV} there is a substantial drop in the number of active cases during this stage. (Right) Scenario 3 with $t_\text{off}=14$, for this configurations there is no second peak after region \textbf{IV}. In both plots we set $t_\text{in}=160$.}
\label{fig:scen3}
\end{figure*}

Scenario \ref{sce:1} is shown in Fig.~ \ref{fig:scen12}. This scenario shows a behavior that is very close to the \emph{official} notified cases\footnote{See the site \url{https://www.boliviasegura.gob.bo} for up to date data.}. In this scenario, the epidemiological peak is expected to occur around day 166 (i.e., around August 22th) assuming that the overall parameters will remain the same until the end of the epidemic. This projection for the peak date also coincides with the range of dates obtained in Sect.~ \ref{sec:fit} and with official authorities information.

\begin{figure}[t]
\includegraphics[width=0.76\textwidth]{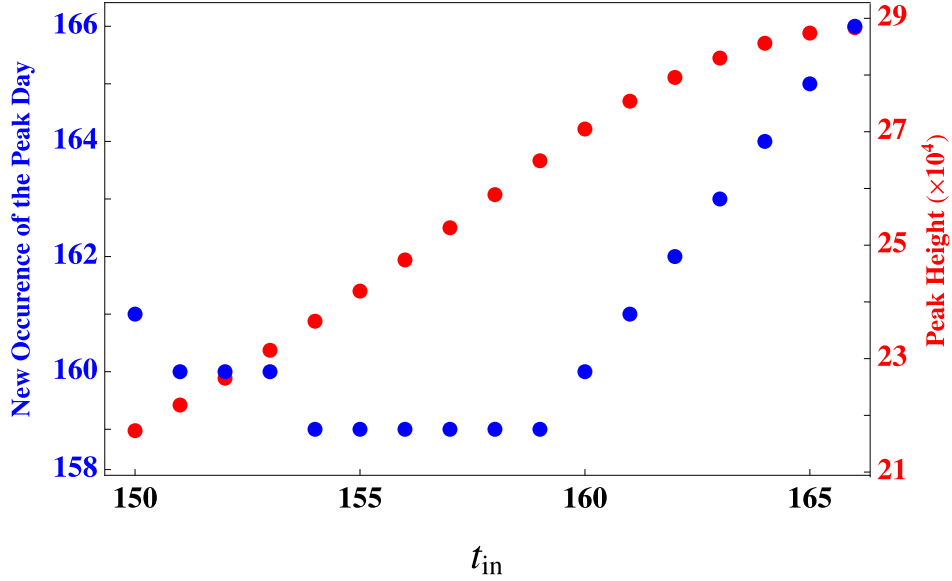}
\caption{Dependence of the peak height and the day of occurrence of this peak with respect to $t_\text{in}$.}
\label{fig:tin}
\end{figure}

Different configurations for $t_{\text{off}}$ in scenario \ref{sce:3} for $t_\text{in}=160$ are shown in Fig.~\ref{fig:scen3}. For the chosen values of $\phi$ for region \textbf{IV} in this scenario, it may be seen that the epidemic peak of scenario \ref{sce:1} is avoided, and the peak may be shifted to occur before the initial predicted date. Fig.~\ref{fig:tin} shows this behavior, by varying the starting date of the rigid quarantine around the peak we show that both the height and the peak day can be lowered by varying $t_\text{in}$. In fact, for the given scenario we show that it is possible to engineer an optimized quarantine. For instance, in Fig.~ \ref{fig:tin} there is an optimal day for the beginning of the quarantine, where the new peak shall occur before the initial predicted day, therefore offering for the possibility of an anticipated economy re-opening, and a considerable reduction in the number of infected people at the peak, allowing for the healthcare system to be able to handle the peak. Fig.~\ref{fig:tin} shows that such value for $t_\text{in}$ is day 152.

\begin{figure*}[b]
\includegraphics[width=0.46\textwidth]{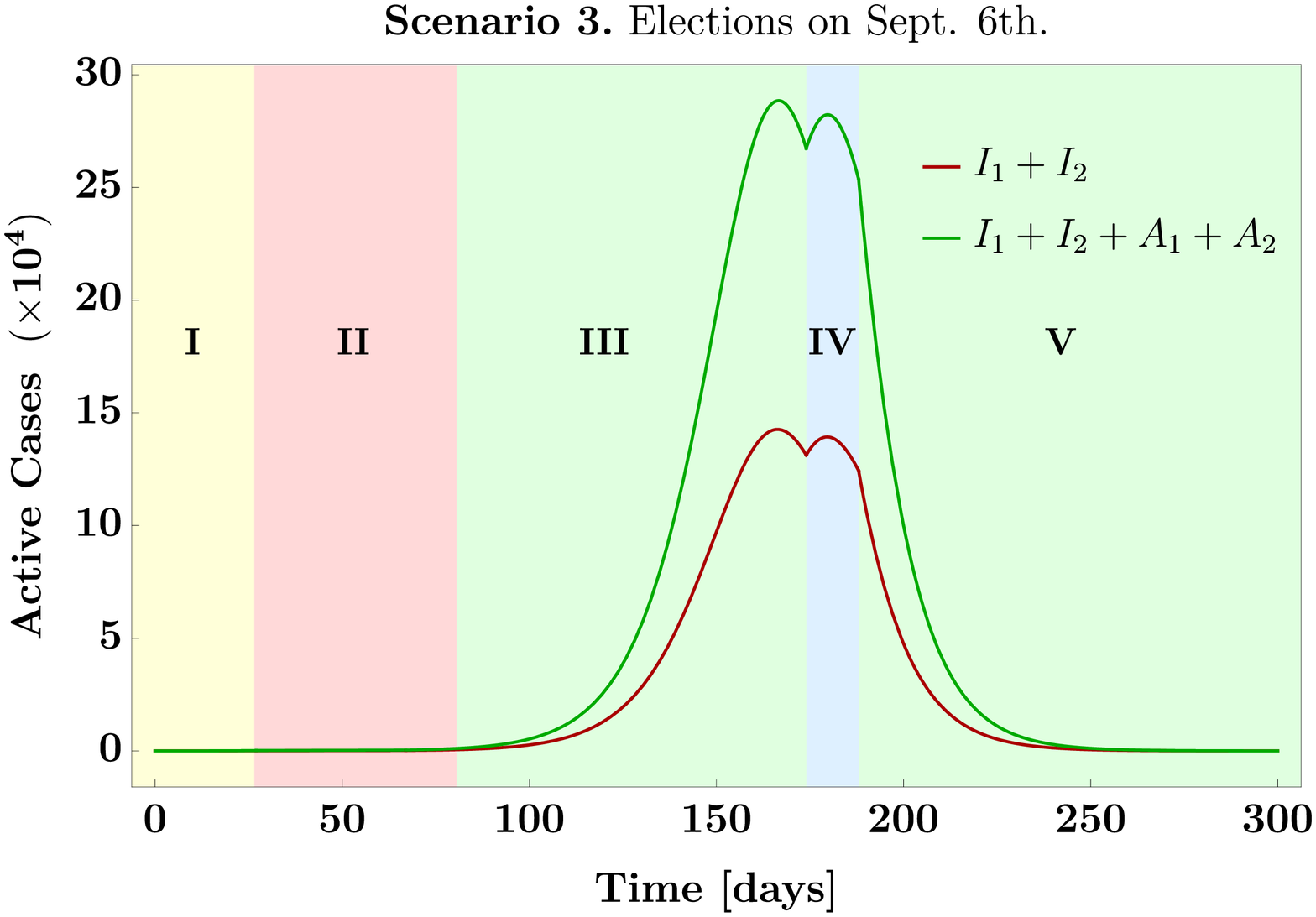}\qquad
\includegraphics[width=0.46\textwidth]{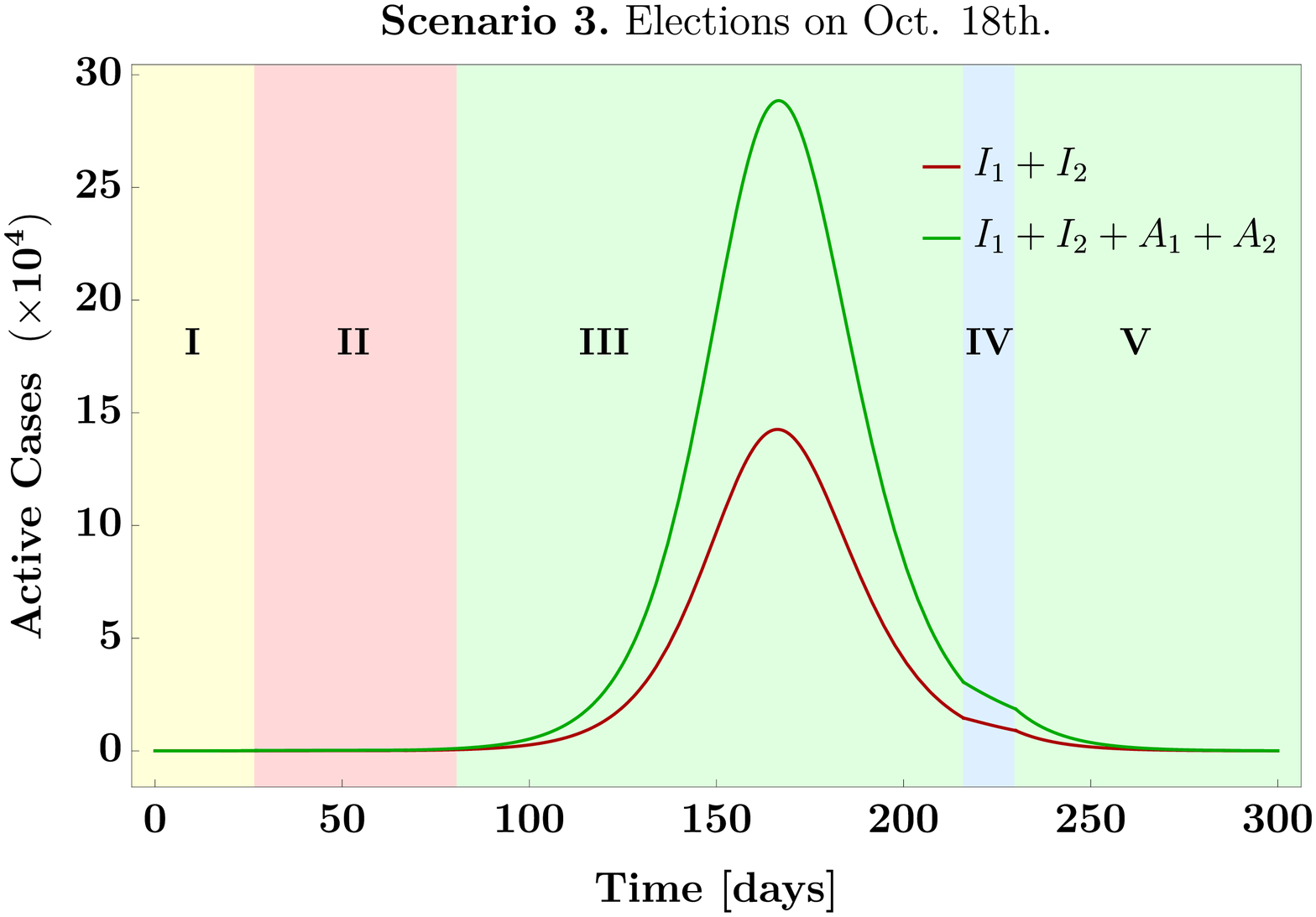}
\caption{(Left) Scenario \ref{sce:5} for the poll day set on September 6th. There exists a second peak due to the massive mobilizations near the elections. (Right) Scenario \ref{sce:5} for the election day set on October 18th. There exists a slight delay in the diminution of the number of active cases due to the elections.}
\label{fig:elecs}
\end{figure*}

The national election scenario is shown in Fig.~ \ref{fig:elecs}. For the case of the poll day occurring on September 6th, the popular mobilizations before and after the democratic event might induce a second peak almost immediately after the first one. This kind of behavior, besides the undeniable fact of the overcharging of a weak healthcare system, can lead to a further uncontrollable situation for the disease's management. Of course, evidence suggests that having a national election immediately after the peak is a reckless action. On the other side, having the vote rescheduled to occur on October 18th\footnote{Based on electoral authorities' information, this is the last possible date to have an election.} could be a less disadvantageous scenario for the potential spreading of the disease. This behavior is mainly due to the fact that the election day is far away from the epidemic peak. Thus, there are less persons in the mobilized  population that can be potential (symptomatic or asymptomatic) carriers of the disease.

\section{Conclusions and Perspectives} \label{sec:conclusions}

The socioeconomic particularities of Bolivian society play an important role in the dynamics of the epidemic, since about $62\%$ of the economy relies on informal work, i.e., on a \emph{day-to-day} based economy. The success rate for any social distancing policy heavily depends on the engagement of the second group at any time of the epidemic.

We have shown that the imposition of a rigid quarantine at any time of the epidemic can affect its dynamics. Furthermore, it is possible to avoid reaching an epidemiological peak by carefully designing a social distancing policy involving the strengthening of all public resources aimed at controlling the epidemic and the overall population's engagement in obeying the social distancing policies. The shift produced in the time at peak occurs as a goal, in principle, to prepare the healthcare system adequately in such a way to avert an over-stressed situation.

We must note that the estimated epidemiological parameters in Section \ref{sec:model} take account of the first 27 days of the epidemic in Bolivia. Social distancing policies were imposed very early in the development of the pandemic, hence, the estimated parameters do not account for an entirely free evolution of the epidemic but instead may consider for the \emph{soft} social distancing imposed in the early stage (closure of schools, reduced working hours) but also for the non-regulated activities during that period (mainly consisting in crowds gathering in public transportation, cultural events, etc.). This fact affects the rest of the evolution of the model presented in Section \ref{sec:model}, since $\phi=1$ at the beginning of the development, this value represents the social scenario of March 2020 and any other scenario with $\phi<1$ describes a situation where social distancing is stricter than that of the early epidemics.

Regarding the electoral scenario, we have shown that the best possible date for the occurrence of the elections corresponds to a time far away from the epidemic peak. In the case of the present work, and based on the estimated parameters (which, for instance, offer reliable projections for July 2020), the peak is expected to occur around mid-August. We emphasize that this scenario can be \emph{drastically} changed due to the social behavior in the midterm. Thus, the peak can easily be shifted for occurring before or after the expected time; and even, it might be lowered.

The vast heterogeneity of environments and the different palliative actions taken in distinct regions of the country imply that each region possesses a particular dynamical scenario for the evolution of the epidemics. However, the numerical evidence shows that the outbreaks' overall behavior can be affected by the imposition of social distancing measures at any time of the epidemic. Of course, it is better to have a rigid quarantine near the peak due to the overcharging of the healthcare system. We highlight that social distancing policy alone cannot be entirely sufficient if other measures do not accompany it. For instance, those involving the insurance that most of the persons will be able to obey the quarantine (especially involving people belonging to the informal economy group), the strengthening of the healthcare system in general including the diagnosis phase, an adequate case tracking and enhancement of treatment resources for the population.

Among further developments of this work, we intend to enhance the epidemiological model to account for different age groups, imposing various social distancing measures to each group. For instance, official data reveals that during the early stage of the epidemic, most of the reported cases (and deceases) consisted of elderly persons. In contrast, as for late July (after having loosened the first rigid quarantine), most of the infected cases account for persons in working age. It could also be important to incorporate in future work the effect of re-contagion. Finally, We consider that the noncompliance degree for each group is an important indicator of social behavior, and it deserves further in-depth analysis.
\section*{Acknowledgments}
MLP acknowledges support from the State Scientific and Innovation Funding Agency of Rio de Janeiro (FAPERJ, Brazil). MLP thanks V. M. Pe\~{n}afiel for useful comments on an early version of this manuscript.

\bibliography{CovBol.bib}
\end{document}